\documentclass[twocolumn,tighten,trackchanges]{aastex62}

\usepackage{amsmath}
\usepackage{natbib}

\newcommand{\code}[1]{\texttt{#1}}
\newcommand{\mesa}{\code{MESA}}
\newcommand{\MESA}{\mesa}

\newcommand{\GYRE}{\code{GYRE}}

\newcommand{\stella}{\code{STELLA}}

\newcommand{\STELLA}{\stella}

\newcommand{\Lsun}{L_\odot}
\newcommand{\Msun}{M_\odot}
\newcommand{\Rsun}{R_\odot}
\newcommand{\Mej}{M_{\rm ej}}
\newcommand{\alpham}{\alpha_{\rm MLT}}

\newcommand{\Eexp}{E_{\rm exp}}

\newcommand{\Lfifty}{L_{50}}

\newcommand{\Lbol}{L_{\rm bol}}

\newcommand{\tpt}{t_{\rm p}}

\newcommand{\Ni}{{^{56}{\rm Ni}}}

\newcommand{\MNi}{M_{\rm Ni}}

\newcommand{\ET}{ET}

\newcommand{\QNi}{Q_{\rm Ni}}

\newcommand{\days}{{\rm d}}

\newcommand{\appropto}{\mathrel{\vcenter{
		\offinterlineskip\halign{\hfil$##$\cr
	\propto\cr\noalign{\kern2pt}\sim\cr\noalign{\kern-2pt}}}}}

\newcommand{\mesafour}{MESA~IV} 

\newlength{\apjcolwidth}
\newlength{\figwidth}
\newlength{\doublewide}
\begin{document}
\title{A Massive Star's Dying Breaths: Pulsating Red Supergiants and Their Resulting Type IIP Supernovae}

\author[0000-0003-1012-3031]{Jared A. Goldberg}
\affiliation{Department of Physics, University of California, Santa Barbara, CA 93106, USA}

\author{Lars Bildsten}
\affiliation{Department of Physics, University of California, Santa Barbara, CA 93106, USA}
\affiliation{Kavli Institute for Theoretical Physics, University of California, Santa Barbara, CA 93106, USA}

\author{Bill Paxton}
\affiliation{Kavli Institute for Theoretical Physics, University of California, Santa Barbara, CA 93106, USA}

\correspondingauthor{J. A. Goldberg}
\email{goldberg@physics.ucsb.edu}

\begin{abstract}
Massive stars undergo fundamental-mode and first-overtone radial
pulsations with periods of 100-1000 days as Red Supergiants (RSGs). 
At large amplitudes, these pulsations substantially modify the outer
envelope's density structure encountered by the outgoing shock wave
from the eventual core collapse of these $M>9M_\odot$ stars.  Using
Modules for Experiments in Stellar Astrophysics (\texttt{MESA}), we
model the effects of fundamental-mode and first-overtone pulsations in
the RSG envelopes,  and the resulting Type IIP supernovae (SNe) using
\texttt{MESA}+\texttt{STELLA}.  We find that, in the case of
fundamental mode pulsations, SN plateau observables such as the
luminosity at day 50, $L_{50}$, time-integrated shock energy $ET$, and
plateau duration $t_{\rm p}$ are consistent with radial scalings
derived considering explosions of non-pulsating stars. Namely, most of
the effect of the pulsation is consistent with the behavior expected
for a star of a different size at the time of explosion. However, in the case of overtone
pulsations, the Lagrangian displacement is not monotonic. Therefore,
in such cases, excessively bright or faint SN emission at different times reflects the
underdense or overdense structure of the emitting region near the
SN photosphere.
\end{abstract}

\section{Introduction\label{sec:INTRODUCTION}}

Periodic variability is prevalent in Red Supergiant (RSG) stars, and is interpreted as 
being a result of radial pulsations \citep{Stothers1969,Stothers1971,Guo2002}.  The mechanism driving 
these pulsations is not fully understood, but they are thought to be driven by
a $\kappa$ mechanism in the hydrogen ionization zone 
with some uncertain feedback within the convective 
envelope \citep{Heger1997, Yoon2010}. \citet{Kiss2006} and \citet{Percy2014} identified periods 
of a few hundred to a few thousand days with varying stellar lightcurve morphology for 
RSGs in the AAVSO International Database. Such pulsations have also been observed occurring in RSGs within
the Small and Large Magellanic Clouds \citep{Feast1980, Ita2004, Szczygiel2010, Yang2011, Yang2012, Yang2018}, M31 and 
M33 \citep{Soraisam2018,Ren2019}, M51 \citep{Conroy2018}, M101 \citep{Jurcevic2000},
within HST archival data of NGC 1326A, NGC 1425, 
and NGC 4548 \citep{Spetsieri2019}, and within the GAIA DR2 RSG sample \citep{Chatys2019}.
These works identify these RSG pulsations as consistent with radial fundamental modes 
and some first radial overtones.

More luminous RSGs generally exhibit longer periods and higher pulsation amplitudes, with 
all RSGs in M31 brighter than $M_k\approx-10$ mag $(\log[L/\Lsun]>4.8)$ varying 
with $\Delta m_R > 0.05\ {\rm mag}$, with R-band variability 
around $\Delta m_R\approx0.4$ in some of the more luminous objects \citep{Soraisam2018}.  
Although it is expected that the metallicity of the host environment might have some small 
impact on the period-luminosity relationship \citep{Guo2002}, this effect is weak compared 
to the scatter within the data (see, e.g. \citealt{Conroy2018, Ren2019, Chatys2019}).
It is not known whether there is a strong relationship between the host metallicity 
and pulsation amplitude, but the amplitudes reported for metal-rich M31 
are similar to the pulsation 
amplitudes of RSGs in M33 
despite the $\approx$0.25 dex
difference in metallicity \citep{Ren2019}. 
There is, however, a noticeable increase in the number ratio of RSGs pulsating in their fundamental 
mode versus the first overtone mode with increasing metallicity \citep{Ren2019}.

Multi-epoch studies of Red Supergiants as potential progenitors for direct collapse into black holes 
are underway \citep{Kochanek2008}, which are ideal for probing the variability of these objects 
as candidates for core-collapse supernovae (CCSNe) as in \citet{Kochanek2017} and \citet{Johnson2018}. 
So far, the majority of supernovae (SNe) whose progenitors have been monitored are consistent with no variability, 
with the exception of the progenitor of the Type IIb SN 2011dh \citep{Kochanek2017}, which was variable in R-band 
by $0.039\pm0.006$ mags per year \citep{Szczygiel2012}. 
This is not inconsistent with the near ubiquity of RSG pulsations at high luminosities,
as most progenitors observed before undergoing Type II SNe have been on the lower end of the RSG 
luminosity spectrum \citep{Smartt2009, Smartt2015}, where pulsation 
amplitudes are likewise generally lower. However, still relatively few such events have been monitored, 
and there is an open theoretical question about how CCSN lightcurves are influenced by 
the presence of progenitor pulsations. 

Recent work highlights that modeling of lightcurves and photospheric velocities  
alone is insufficient to extract progenitor characteristics from observed SNe \citep{Dessart2019, 
Goldberg2019, Martinez2019}. 
A progenitor radius can provide a crucial constraint, allowing to distinguish
between, say, a more compact higher ejecta-mass event with a higher explosion energy, 
and an event with a larger progenitor radius, lower ejecta mass, and lower explosion energy. 
This has been done recently by creating matching lightcurve models for SNe with observed 
progenitor radii (e.g. \citealt{Martinez2019}), fixing a mass-radius relationship by 
fixing stellar evolution parameters (such as metallicity, mixing length in 
the H-rich envelope, overshooting, winds) and fitting to a large set of population
synthesis lightcurve models (e.g. \citealt{Eldridge2019}), and in an ensemble fashion by using a 
prior on the radius of RSGs to extract explosion energies statistically for an existing 
sample of IIP lightcurves \citep{Murphy2019}. Because, in reality, the progenitor radius 
could be affected by RSG pulsations, this could lend itself to additional uncertainty 
in any explosion parameters recovered from SN observations, 
especially in the case of directly using an observed progenitor radius at an unknown phase 
relative to the time of explosion. 

Observed Type IIP SNe are also often reported to show excess emission before day $\approx$30,
often attributed to interaction with the extended environment surrounding the progenitor (e.g. 
\citealt{Khazov2016, Morozova2017, Morozova2018, Forster2018, Hosseinzadeh2018}). 
Because models of early emission depend sensitively on the progenitor density profile 
(e.g. \citealt{Nakar2010, Sapir2011, Katz2012, Sapir2017, Faran2019}), any modification 
of the outer stellar structure and surrounding environment could translate to distinct 
changes in the early SN emission (see, e.g., \citealt{Morozova2016}). 
For example, the effects of pulsation-driven superwinds \citep{Yoon2010} on early SN-IIP lightcurves 
have been directly considered by \citet{Moriya2011,Moriya2017}. 
However, 1D modeling of the extended atmospheres of massive stars
is inherently limited, as 1D codes cannot reproduce the detailed 3D structure of 
the outermost envelope (see e.g. \citealt{Chiavassa2011,ArroyoTorres2015,Kravchenko2019}). 
Therefore, in this work we primarily restrict our discussion to plateau properties 
after day $\approx$30, at which point the SN emission comes from the modified interior of 
the star and not the outermost $\approx0.2\Msun$.

In this work, we consider effects of pulsations on the bulk density structure of 
the stellar envelope and the impact these structural differences have on the 
resulting Type IIP SNe. In Section \ref{sec:PULSATIONS} we discuss our 
approach to capturing the effects of radial pulsations on the internal structure 
of the star using the open-knowledge 1D stellar evolution software instrument 
Modules for Experiments in Stellar Astrophysics (\MESA; 
\citealt{Paxton2011,Paxton2013,Paxton2015,Paxton2018,Paxton2019}), 
and compare our pulsating models to expectations from linear theory. 
In Section \ref{sec:EXPLOSIONS} we demonstrate the effects these structural 
changes have on the resulting SN lightcurves. We show the luminosity at day 50 
($\Lfifty$), time-integrated shock energy ($\ET$), and plateau duration ($\tpt$)
for SNe of progenitors pulsating in their fundamental mode scale simply with 
the progenitor radius at the moment of explosion as given by 
\citet{Popov1993, Kasen2009, Nakar2016, Goldberg2019} and others. 
Furthermore, we show that for pulsations where the displacement is 
not monotonic, such as the first overtone, SN emission from different regions 
within the ejecta is influenced by the differing structure.

\section{Modeling Radial Pulsations\label{sec:PULSATIONS}}

We construct our fiducial model of a CCSN progenitor with \MESA\ revision 11701. 
We choose a nonrotating, solar-metallicity ($Z=0.02$) model of 18$\Msun$ at ZAMS, with a convective efficiency of $\alpham=3.0$ 
in the Hydrogen-rich envelope. We use modest convective overshooting parameters $f_{\rm ov}=0.01$ and $f_{\rm 0,ov}=0.004$, 
and winds following \MESA's `Dutch' prescription with efficiency $\eta_{\rm wind}=0.4$ \citep{DUTCH,DUTCH2,DUTCH3}.  
After the end of core carbon burning, identified when the central fraction of $^{12}$C falls below $10^{-6}$, 
we introduce a maximum timestep of $10^{-3}$ years. This is to ensure that the model remains numerically 
converged, as well as to ensure that we resolve changes its structure when causing it to pulsate on a timescale of 
hundreds of days. Other inputs are determined following the \verb|25M_pre_ms_to_core_collapse| case of the \mesa\ test suite. 
At the time of core-collapse, 1715 days after the end of core Carbon burning, the unperturbed model 
has a total mass of $M=16.3\Msun$, a radius of $R = 880\Rsun$, and a luminosity of $L = 1.56\times10^5\Lsun$. 

\begin{figure}
\centering
\includegraphics{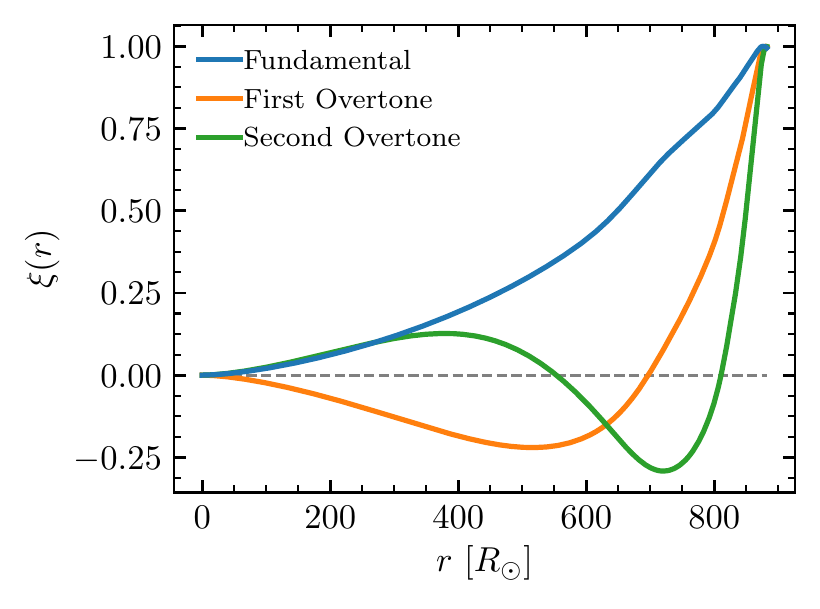} 
\caption{Normalized radial displacement eigenfunctions for our fiducial stellar model at core Carbon depletion.}
\label{fig:eigenfunction}
\end{figure}

After evolving the model through the end of core carbon burning, we use the pulsation instrument \GYRE\ 
\citep{townsend_2013_aa} to identify the periods and radial displacement eigenfunctions for the first 
3 radial ($l=0$) modes. We recover a fundamental pulsation period of 534 days, a first overtone period 
of 240 days, and a second overtone period of 154 days. The radial displacement eigenfunction $\xi(r)$ for the 
fundamental mode, and the first and second overtones, normalized to $\mathrm{max}(\xi(r))=1$, 
are shown in Figure \ref{fig:eigenfunction}. 

\begin{figure}
\centering
\includegraphics[width=\columnwidth]{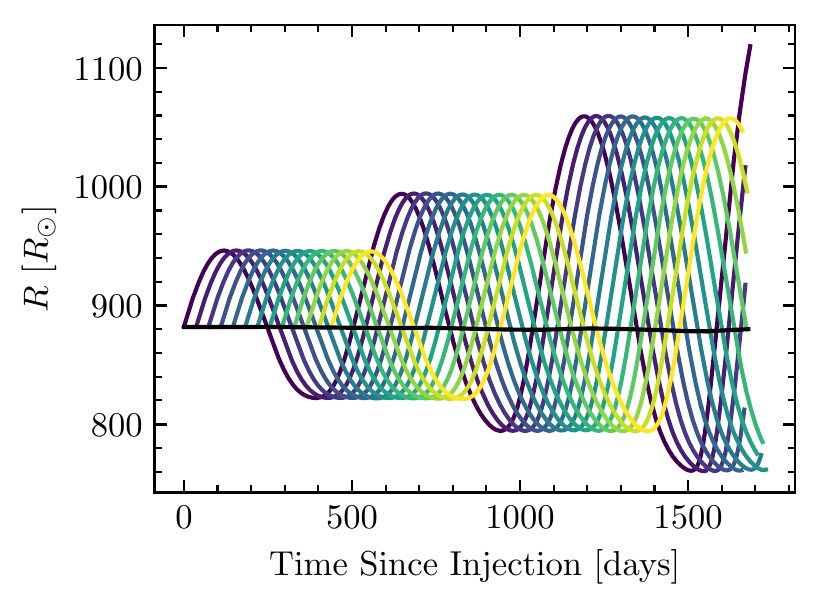} 
\caption{Stellar radius as a function of time, after injecting the velocity eigenfunction of the fundamental 
radial mode. The left-most point on each curve corresponds to the time of injection relative to the 
earliest injection, and the right-most point corresponds to the model at the time of core collapse. 
The black line shows the negligible variation in the stellar radius of the unperturbed model.}
\label{fig:ringing}
\end{figure}

To model the effects of pulsation on the density structure of the envelope, we inject the fundamental eigenmode 
as a velocity proportional to the radial displacement given by \GYRE. For a zone with radial 
coordinate $r$, we set $v(r)= 1.2\,c_{\rm s,surf}\,\xi(r)$, where $c_{\rm s,surf}$ is the sound speed 
at the surface of the unperturbed model and $\xi(r)$ is normalized to be 1 at its maximum value. 
The resulting pulsation causes significant variation in the radius, from 760 - 1100 $\Rsun$ over the course of a few pulsations. 
This amplitude was chosen to resemble the 0.3-0.4 mag amplitudes seen by \citet{Soraisam2018}.
We do not claim that the growth in the pulsations is being modeled correctly; rather,  
we are only interested in the effects of realistically large pulsations on the SN properties.
In order to achieve core collapse at different phases of the pulsation, we inject this velocity 
eigenfunction starting at increments of 36.5 days up to 474.5 days after core carbon depletion and allow the model to ring 
as it evolves to core collapse, as shown in Figure \ref{fig:ringing}. For the fundamental mode, 
the recovered average peak-to-peak period is 535 days, and trough-to-trough period is 550 days, as the pulsation becomes
increasingly nonlinear, especially near the minimum radius. However, both are close to the 534 day period expected of a 
small amplitude pulsation.

\begin{figure}
\centering
\includegraphics[width=\columnwidth]{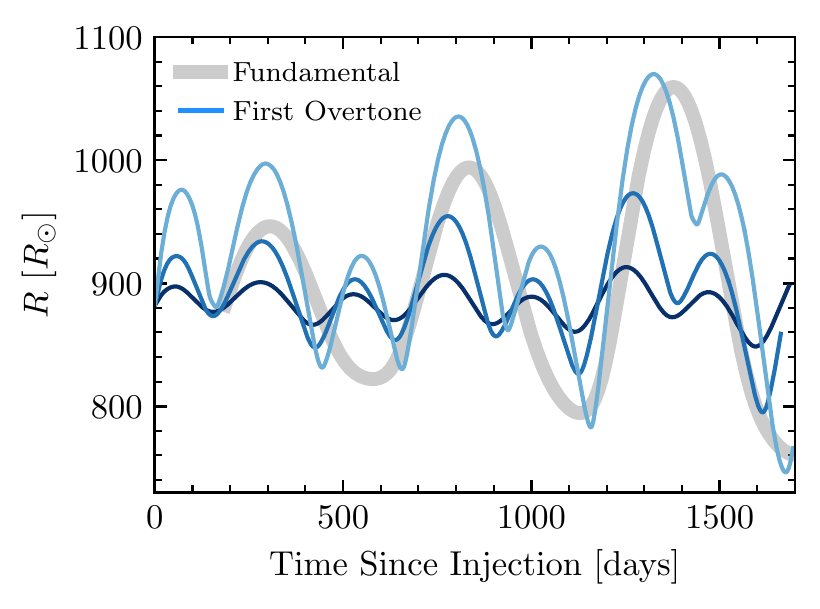} 
\caption{
Stellar radius as a function of time in our models injected with first overtone velocity 
eigenfunctions. The injected initial velocity amplitudes shown here are $A=0.69$ (dark blue), 
1.71 (average blue), and 3.42 (light blue), for 
velocities injected of the form $v(r)= A\,c_{\rm s,surf}\,\xi(r)$ where $\xi$ is the displacement 
eigenfunction for the first overtone. A fundamental mode pulsation is also shown, with its starting 
point chosen to visually resemble the modulation seen in the overtone models.
}
\label{fig:overtone_diff_amps}
\end{figure}

The process of causing our models to pulsate with the first radial harmonic is nearly identical to 
that described above. However, since the overtone pulsation period of 240 days 
is approximately half that of the fundamental mode, and there is a node in the 
radial displacement eigenfunction such that the surface displacement is only caused by oscillation in the 
outer envelope, the radial pulsation amplitude is comparatively small for a given injected velocity amplitude. 
Figure \ref{fig:overtone_diff_amps} shows the overtone pulsation injected with different amplitudes. A fundamental
mode is also shown for comparison. 
The recovered average peak-to-peak and trough-to-trough periods are 236 days and 241 days, respectively, taken 
over the first 4 pulsation cycles. Particularly for larger amplitude pulsations, 
the fundamental mode grows in the overtone-injected models, causing modulation on longer timescales than 
the overtone period. This effect gets stronger with increasing initial pulsation amplitude, making it very difficult 
to create a model which rings with a ``pure" overtone and has a sizeable pulsation amplitude. 

\subsection{Analytic Expectations in the Linear Regime}

\begin{figure}
\centering
\includegraphics[width=\columnwidth]{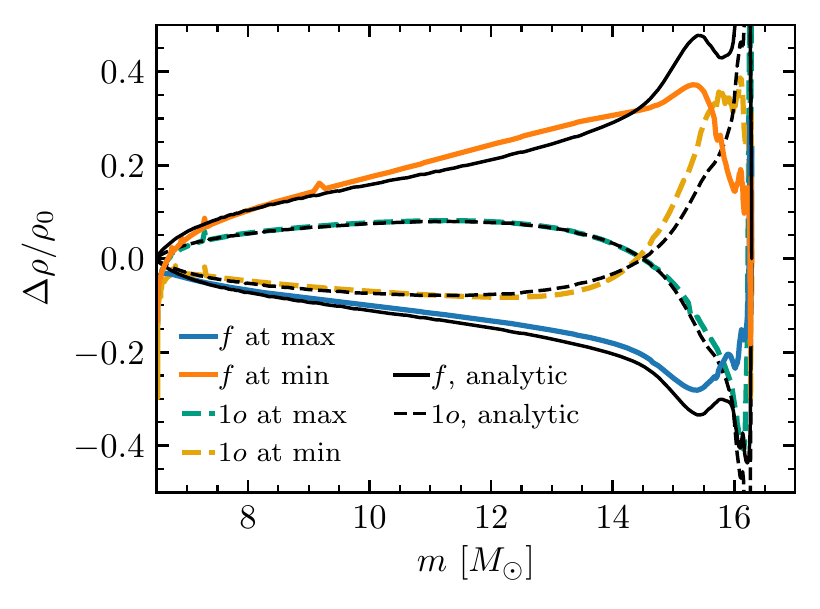} 
\caption{Comparison of linear theory for the Lagrangian density perturbation (black lines) 
with differences in the model density profiles from the density profile of the unperturbed 
starting model (colored lines) 
for fundamental mode pulsations (solid) and first overtone pulsations (dashed).}
\label{fig:LagrangeDensity}
\end{figure}

For a small perturbation, we can express the radius of that element as
$\vec{r} = \vec{r}_0 + \vec{\xi}$, where $\vec{r}_0$ is the unperturbed radius and $\vec{\xi}$ is the Lagrangian displacement.
For a radial oscillation with $\vec{\xi} = \xi e^{i\omega t}\,\hat{r}$, where $\omega$ is the frequency of oscillation,
the velocity of that fluid element is $\vec{v} = i \omega \vec{\xi}$. By continuity, the density of the fluid element changes as
\begin{equation}
\frac{d\rho}{d t} + \rho \vec{\nabla}\cdot\vec{v} = 0,
\label{eq:continuity}
\end{equation}
where $d/dt$ represents the Lagrangian time derivative $d/dt = \partial/\partial t + \vec{v}\cdot\vec{\nabla}$. 
Equation \eqref{eq:continuity} yields the Lagrangian density perturbation $\Delta\rho$,
\begin{equation}
\Delta\rho = -\rho_0 \vec{\nabla} \cdot \vec{\xi} = -\rho_0 \frac{1}{r^2}\frac{d}{dr}r^2\xi.
\label{eq:lagrangianrho}
\end{equation}
In order to check the agreement between our pulsating model and the expectations 
from linear theory, we save the density profile at the maximum and minimum radius for fundamental mode 
and overtone pulsations. Figure \ref{fig:LagrangeDensity} shows the agreement between our models 
and Equation \eqref{eq:lagrangianrho}. Here we normalize $\xi$ to match the displacement in the 
pulsating model at the mass coordinate corresponding to $300\Rsun$ in the unperturbed model, at an 
overhead mass of 5.7$\Msun$. This location was chosen because it corresponds to roughly half of the 
envelope mass and half of the stellar radius in log-space. The surface is most severely affected by 
nonlinearities, and this work primarily explores effects on the bulk of the material.
We also choose to display the overtone profiles at the first maximum (1/4 period after injecting the 
velocity eigenfunction) and the second minimum (7/4 period after injection) of the model with an injected velocity of 
$v(r)= 1.71\,c_{\rm s,surf}\,\xi(r)$, as these times are most consistent with being ``pure" overtones.
The agreement is very good in the interior of the star.
Deviations from linear theory occur primarily near the surface,
where nonlinearities due to nearly sonic motion cause a larger impact.

\section{Exploding Pulsating Models\label{sec:EXPLOSIONS}}

\begin{figure*}
\centering
\includegraphics[width=\columnwidth]{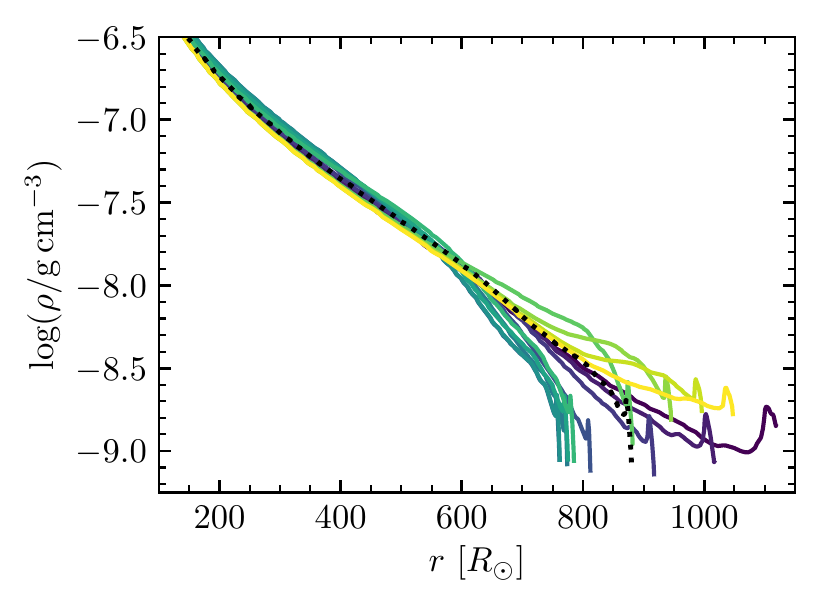} 
\includegraphics[width=\columnwidth]{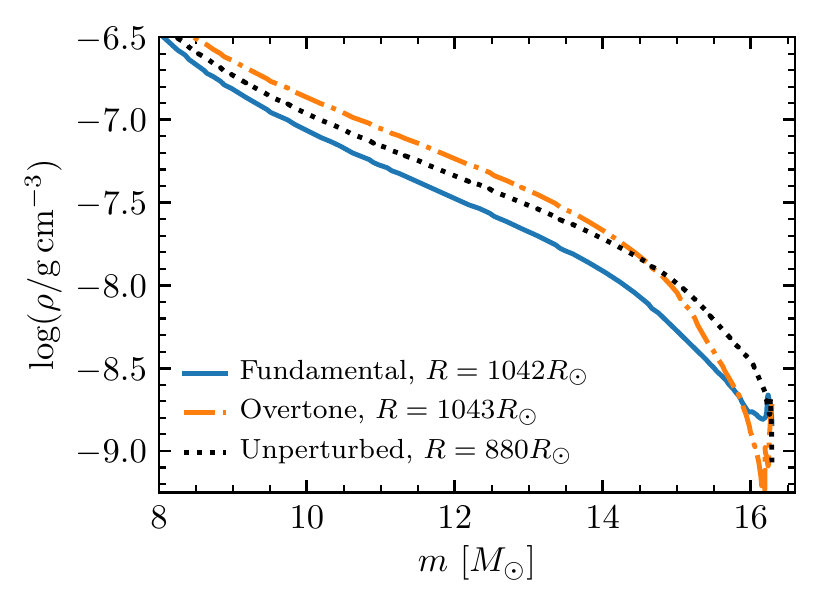}
\caption{Left: Density profiles in the envelope of our pulsating models just before core collapse, where 
color corresponds to time the pulsation was injected as in Figure \ref{fig:ringing}. 
Right: Lagrangian density profiles at core-collapse for 
large-amplitude pulsations approaching maximum displacement, where the velocity eigenfunctions
were injected just 1/4 phase before core-collapse to preserve the purity of the modes. In both panels, the 
dotted black line shows the unperturbed model.}
\label{fig:density_profiles}
\end{figure*}

At the time of explosion, the density profiles in the envelope 
vary significantly for different pulsation phases. This can be seen in Figure \ref{fig:density_profiles}, 
which shows density profiles in the envelope at core-collapse for the fundamental-mode models 
as a function of radius (left panel). 
Additionally, Figure \ref{fig:density_profiles} 
shows a comparison between Lagrangian density profiles of the unperturbed model, a fundamental mode pulsation 
near maximum, and a large-amplitude overtone near maximum (right panel). In order to achieve 
a large-amplitude overtone pulsations, we inject a velocity profile 
with $v(r)= 5.48\,c_{\rm s,surf}\,\xi(r)$, where $\xi$ is the displacement for the first overtone,
approximately quarter-period before core-collapse, 1533 days after core C depletion, so that it is approaching its
first maximum at the time of explosion. To produce a fundamental mode pulsator with the same stellar 
radius and similar phase, we inject a velocity profile $v(r)= 2.86\,c_{\rm s,surf}\,\xi(r)$ approximately 
a quarter-period before core-collapse, 1460 days after core C depletion. 
Our models show significant diversity in their density profiles, particularly near the surface. 
Moreover, the overtone pulsation at maximum phase is denser in the interior of the star compared to the 
unperturbed model, but less dense near the surface, whereas the fundamental mode near maximum 
is less dense everywhere. 

We explode our models at different radii. 
At a central temperature of $\log(T_c/{\rm K})=9.9$, we instantaneously zero 
out the velocity profile to ``freeze in" the density structure of the envelope, since the time to shock breakout 
($\approx2$ days)
is much shorter than the pulsation period, and since the kinetic energy associated with the pulsation 
is orders of magnitude below the total binding energy of the star. This also helps quell artificial velocity 
fluctuations which begin to arise in the core around the time of core Si burning. We then continue to evolve the model until 
core infall. At that point, we excise the core, as described in section 6.1 of \citet{Paxton2018} 
(hereafter \mesafour). 
Because each model is evolved independently after core C burning, there is some small variation in the 
excised mass, ranging from 1.6 to 1.74$\Msun$, 
leading to ejecta masses of $\Mej=$14.54 to 14.68$\Msun$. 
The unperturbed model has an excised mass of 1.73$\Msun$.
We allow the new inner boundary to infall until it reaches an inner radius of 500 km. 
We then halt the infall, and inject energy in the innermost 0.1 $\Msun$ of the star for $10^{-4}$ seconds, 
until each model reaches a total energy of $10^{51}$ ergs. 

We proceed by modeling the evolution of the shock including 
Duffell RTI \citep{Duffell2016}, and hand off the ejecta model at shock breakout to the 1D radiation-hydrodynamics 
software \STELLA\ \citep{Blinnikov1998, Blinnikov2000, Blinnikov2006,Baklanov2005} as described in \mesafour. The time to shock breakout is 2 days for the unperturbed model, 
and varies from 1.7 days for our smallest-radius model to 2.5 days for our largest-radius model. 
At this explosion energy, there is negligible additional fallback, which 
we evaluate using the fallback scheme described in Appendix A of \citet{Goldberg2019} with an additional velocity 
cut of 500 km s$^{-1}$ at handoff to \stella. 
We then rescale the distribution of $\Ni$ to match a total mass of 0.06 $\Msun$, which is typical 
of observed events and roughly matches the Ni masses observed in SNe with $\Lfifty$ equal to that of the 
unperturbed model via the $\Lfifty-\MNi$ relations from \citet{Pejcha2015a} and \citet{Muller2017}. 
We use 1600 spatial zones and 40 frequency bins in \STELLA, which yields convergence in the bolometric lightcurves
for the given ejecta models (see also Figure 30 of \mesafour\ and the surrounding discussion). 
While a significant fraction of SNe II-P have excess emission for the first $\sim20$ days (e.g. \citealt{Morozova2017}), 
and pulsation-driven outbursts have been proposed as one means of mass loss at the end of the lives of RSGs
(e.g. \citealt{Yoon2010}), 
we do not include any extra material beyond the progenitor photosphere to generate our model lightcurves. 
In addition, we are focused on the emission from 
the bulk of the ejecta, that occurs after day 30.

\subsection{Pulsations and Plateau Properties}

\begin{figure}
\centering
\includegraphics[width=\columnwidth]{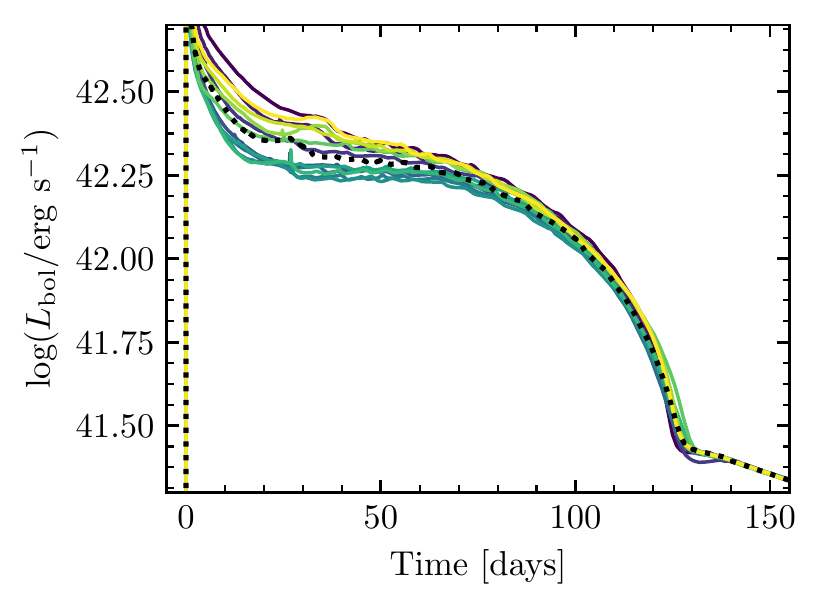} 
\caption{Lightcurves for our fundamental mode pulsator at different phases of pulsation. Color corresponds 
to time the pulsation was injected, as in Figure \ref{fig:ringing}, and tracks pulsation phase. 
The dotted black line shows the lightcurve of the unperturbed model.}
\label{fig:lightcurves}
\end{figure}

\begin{figure}
\centering
\includegraphics[width=\columnwidth]{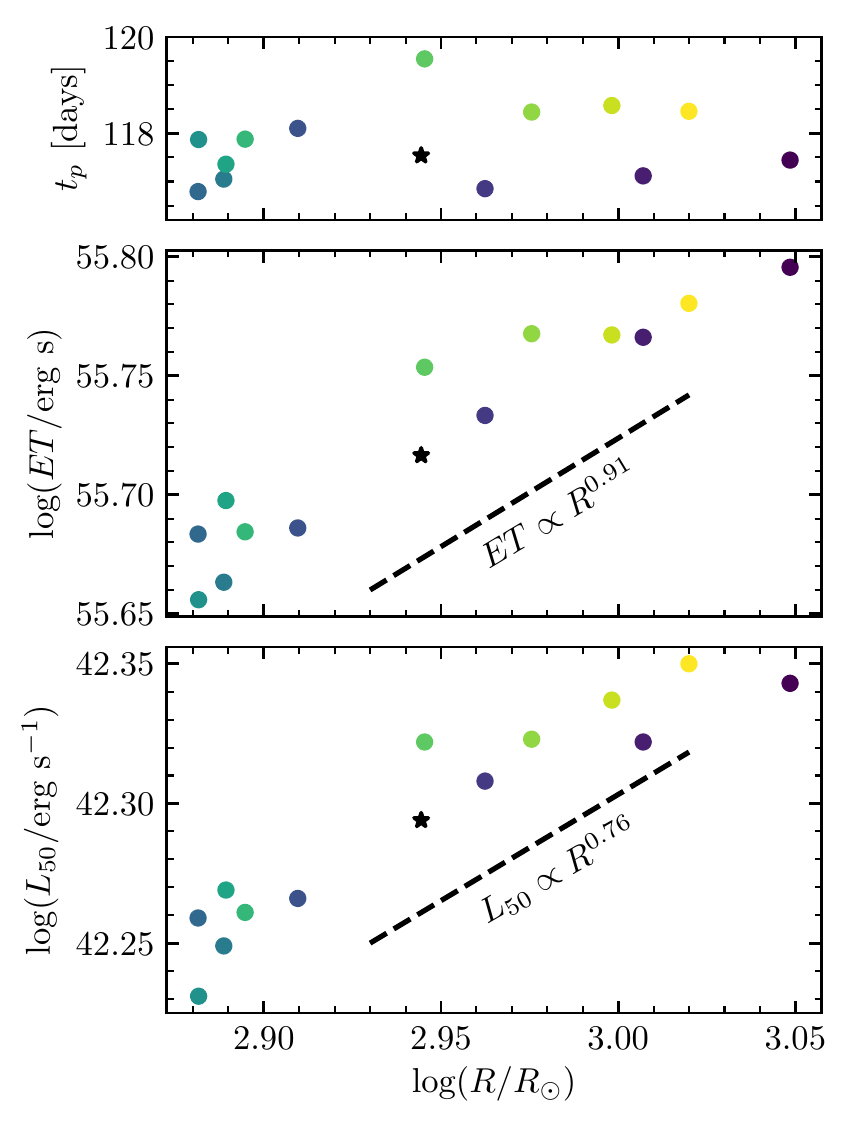} 
\caption{
Lightcurve observables versus progenitor radius at the time of explosion 
for our unperturbed model (black star) and pulsating models (colored points). 
The plateau duration (upper panel), $\ET$ (middle panel), and $\Lfifty$ 
(lower panel) are shown along with scalings from \citet{Goldberg2019}. 
Colors match the colors in Figures \ref{fig:ringing} and \ref{fig:lightcurves}. 
}
\label{fig:SCALINGS}
\end{figure}

As discussed in detail by \citet{Arnett1980}, \citet{Popov1993}, \citet{Kasen2009}, 
\citet{Sukhbold2016}, \citet{Goldberg2019}, and others, the plateau luminosity of a Type IIP SN
at day 50, $\Lfifty$, depends on the radius of the progenitor. 
\citet{Popov1993} gives
$\Lfifty\propto R^{5/6}$
at fixed ejecta mass $\Mej$ and explosion energy $\Eexp$. 
From a suite of \MESA+\stella\ models, 
\citet{Goldberg2019} recovered a similar scaling, $\Lfifty\propto R^{0.76}.$
Figure \ref{fig:lightcurves} shows lightcurves for the 13 phases of pulsation 
shown in Figure \ref{fig:ringing}, as well as for the unperturbed model 
denoted by the black line in Figure \ref{fig:ringing}. As expected from 
the scalings, the luminosity at day 50 varies by 0.13 dex, or 0.33 mag, 
with the brighter explosions corresponding to larger radii, with radii 
ranging from 760-1120$\Rsun$.  The slope on the plateau 
is somewhat steeper in the brighter SNe, such that the variation at early
time is greater than closer to the end of the plateau.
Additionally, following \citet{Goldberg2019}, in the $\Ni$-rich limit
$\MNi\gtrsim 0.03\Msun$, the plateau duration should be approximately independent 
of the progenitor radius, with some variation for varied distributions of $\Ni$ 
and Hydrogen. This can also be seen in our lightcurves in Figure 
\ref{fig:lightcurves}, where the recovered plateau durations (using 
the method of \citealt{Valenti2016} as in \citealt{Goldberg2019}) ranges 
from 116.8 to 119.5 days with no correlation with progenitor radius. 
These trends are shown in greater detail in the upper and lower 
panel of Figure \ref{fig:SCALINGS}, which show good agreement between our 
models and the scalings. 

Figure \ref{fig:density_profiles} also shows changes in the outer density 
profiles and their slopes as a result of these pulsations. These changes do 
modify the calculated early lightcurves shown in Figure \ref{fig:lightcurves}, 
causing greater luminoisty excesses at early times in the more extended models. 
In observations, such apparent excesses are often interpreted as evidence 
for material beyond the normal stellar photosphere. However, because this part 
of the outer envelope is intrinsically uncertain in 1D models, we are not 
in a position to make strong claims about whether the variety seen in early 
lightcurve observations can be explained by pulsations alone. 

Additionally, the total energy deposited by the shock is reflected in the observable 
$\ET$ \citep{Nakar2016, Shussman2016}, defined as the total time-weighted energy radiated 
away in the SN which was generated by the initial shock and not by $\Ni$ decay:
\begin{equation}
\ET=\int_{0}^{\infty} t \left[\Lbol(t) - \QNi(t)\right]\,\mathrm{d} t,
\label{eq:ETdef}
\end{equation}
where $t$ is the time in days since the explosion and
\begin{equation}
\QNi=\frac{\MNi}{\Msun}\left(6.45e^{-t/8.8\days} + 1.45e^{-t/113\days}\right)\times10^{43}\ {\rm erg\ s^{-1}},
\label{eq:QNi}
\end{equation}
is the $\Ni$ decay luminosity given in \citet{Nadyozhin1994}, 
which is taken to be equivalent to the instantaneous heating rate of the ejecta 
assuming complete trapping. 
$\ET$ also scales with the progenitor radius for constant $\Mej$ and $\Eexp$, given as
$\ET\propto R$
by the analytics and modeling of 
\citet{Nakar2016, Shussman2016, Kozyreva2018}, and as
$\ET\propto R^{0.91}$
recovered from \MESA+\stella\ models by \citet{Goldberg2019}.
The middle panel of Figure \ref{fig:SCALINGS} shows the agreement between 
$\ET$ in our model lightcurves and the scalings. Like with $\Lfifty$, 
$ET$ as a function of progenitor radius exhibits some scatter, which is not surprising 
given the significant differences in the density profiles especially 
in the models near pulsation minima at core-collapse, but overall agrees well with 
the predicted scalings. 

\subsection{Comparing Fundamental and Overtone Pulsations}
Although a majority of observed pulsating RSGs are dominated by the fundamental mode, there is evidence
for some pulsating with the first overtone (e.g. \citealt{Kiss2006, Soraisam2018, Ren2019}). 
Because of the radial crossing in the overtone, the progenitor radius used in scaling laws may not be sufficient 
to predict $\Lfifty$.  
Typically, the expansion time characterized by the time to shock breakout 
and the mean density of the SN ejecta are considered in analytics. However, 
the local radius and density profile of the progenitor at the mass coordinate of the SN photosphere, which is located
near the H-recombination front and is defined by the location where the mean optical depth $\tau=2/3$, must be
taken into account. As seen in the left panel of Figure \ref{fig:density_profiles}, inside the mass coordinate
of $\approx14.5-15\Msun$, which is near the zero-crossing in the radial displacement 
(see Figure \ref{fig:LagrangeDensity}), 
the overtone progenitor model is denser than the unperturbed model, and outside that coordinate it is less dense. 
On the other hand, the fundamental mode pulsation is less dense everywhere when it is at a positive 
radial displacement, suggesting that at fixed photospheric mass coordinate in the SN, 
the star should appear ``larger" and therefore the SN would be brighter.

\begin{figure}
\centering
\includegraphics[width=\columnwidth]{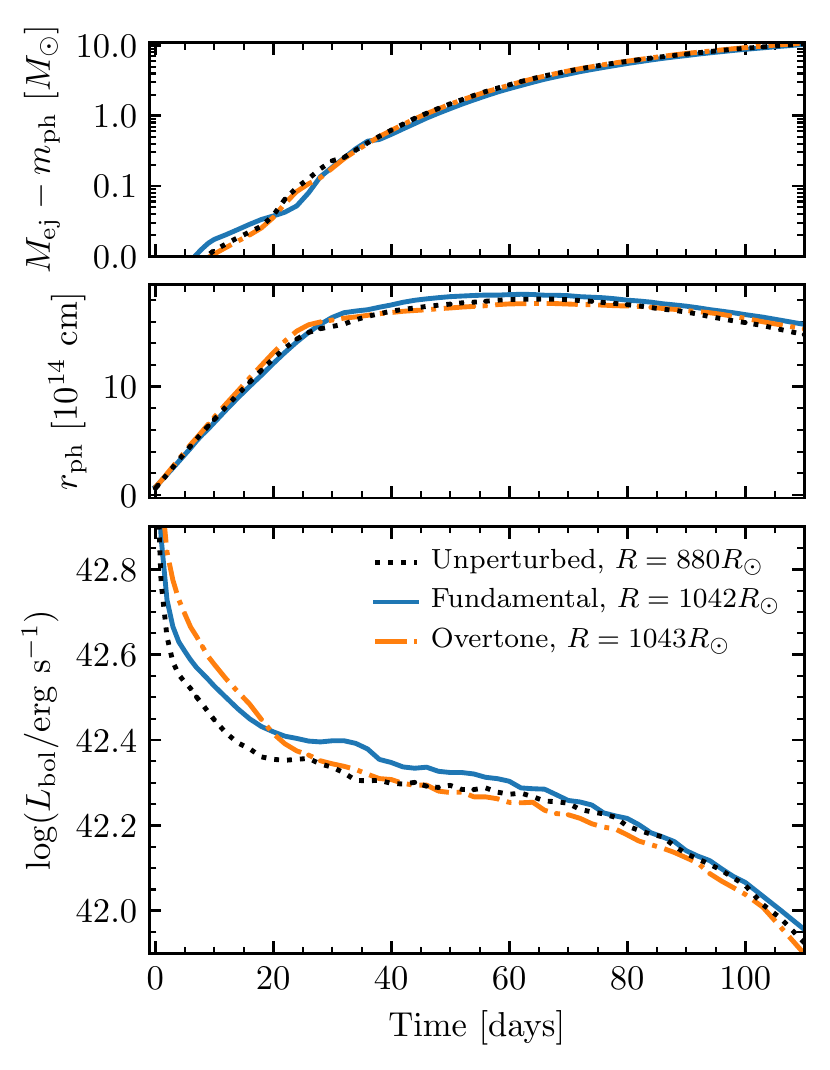} 
\caption{Overhead mass coordinate of the SN photosphere (upper panel), photospheric radius (middle panel), and lightcurves
(lower panel) for explosions of large-amplitude fundamental mode and overtone pulsations near maximum, 
compared to the unperturbed model.}
\label{fig:overtone_compare}
\end{figure}

As shown in the upper panel of Figure \ref{fig:overtone_compare}, the evolution of the
mass coordinate of the SN photosphere does not change significantly for the pulsating models compared to the 
unperturbed model. At day 50, the SN photosphere has moved $1.5\Msun$ into the ejecta for the unperturbed and 
overtone models, corresponding to a stellar mass coordinate of $14.8\Msun$, which is 
near the zero-crossing in the overtone displacement and density perturbation in the progenitor model. 
This is reflected by the lightcurves shown in the lower panel of Figure \ref{fig:overtone_compare}. 
The evolution of the photospheric radius (middle panel of Figure \ref{fig:overtone_compare}) and mass coordinate  
do not differ tremendously on the plateau between the three models, but the lightcurves
show a distinct difference. Whereas the progenitor radii for the fundamental and overtone are nearly identical, 
the overtone explosion at day 50 is fainter by 0.046 dex or 0.115 mags, and in fact much closer in $L_{50}$ to the
unperturbed progenitor model than to the fundamental mode. 
Additionally, the SN from the overtone pulsator is brighter at early times, 
when the SN emission is coming from what appears to be 
a more radially extended star with a steeper density profile, and fainter at later times, when the emission 
appears to be coming from a more compact star.

\section{Conclusions}
There is strong observational evidence for variability in large samples of RSGs caused by radial pulsations
in their envelopes, typically with periods between a few hundred and a few thousand days 
\citep{Kiss2006,Soraisam2018,Chatys2019}. 
Since the final stages of burning take place over week-long timescales, much shorter than the pulsation period, 
the density structure of the envelope can reflect any pulsation phase at the time of explosion. 
This is significant, as the radius and density structure of
a given Type IIP SN progenitor are important in determining the luminosity evolution of its resulting SN. 

We consider the effects of pulsations on the stellar envelope and SN emission after core-collapse. 
We show that SNe of fundamental mode pulsators, which account for the majority of
observed pulsating RSGs, behave like ``normal" Type IIP SNe from progenitors at different radii. 
We find that $\Lfifty$ and $\ET$ scale with the progenitor radius at the time of explosion 
consistent with the work of \citet{Popov1993, Kasen2009, Nakar2016, Goldberg2019} and others, 
and that the plateau duration remains independent of progenitor radius as expected in the $\Ni-$rich regime. 
The luminosity plateau declines more steeply for brighter events between days 30 and 80, 
which in this study correspond to models with positive 
radial displacement at the time of core collapse. This is consistent with the observed 
correlation seen in Type II SNe more broadly between the brightness and steeper plateau decline 
(e.g. \citealt{Anderson2014,Valenti2016}). 

Additionally, we show that large-amplitude pulsations in the first overtone yield different lightcurves
compared to fundamental-mode pulsations at the same radius. This results from the nonmonotonic overtone density 
perturbation, which, for an explosion near pulsation maximum, causes the SN to ``see" a puffier star at early times, 
but a more compact star at later times. This yields a supernova which is initially brighter than 
either the fundamental-mode pulsator at equivalent radius or the unperturbed model at a smaller radius, 
but fainter once emission is coming from the denser interior. In all cases, 
the differing stellar radii and density profiles also yield signatures in the calculated 
early SN emission, but future work aided by a more accurate treatment of the progenitor's extended 
atmosphere is necessary to make definitive statements and quantitative predictions.

Motivated by the observed oscilllations, we 
only considered the impact of radial pulsations on the 
resulting SNe light curves. Non-radial pulsations, if present, 
would lead to additional phenomena, for example apparent asymetries 
during the plateau phase. Existing spectropolarimetric observations 
(\citealt{Wang2001,Leonard2001a,Leonard2001b,Leonard2006,Wang2008,Kumar2016,Nagao2019})
sometimes show very low (or undetectable) levels of assymetries during the plateau, 
with increasing polarization evident in the late time tail 
attributed to asymmetries deep in the helium core. 

Because a fundamental uncertainty in recovered explosion properties from Type IIP SNe stems 
from the unknown radius at the time of core-collapse, the presence of a pulsation would translate to an 
additional uncertainty in recovering progenitor properties 
from SN lightcurves even in conjunction with progenitor detections. 
Therefore, continued studies of RSG variability will be important in determining the 
uncertainties within a single progenitor radius detection. 
Future work is also needed to accurately model the winds and surface layers of 
massive stars, as well as the density profile of any extended material, all of which 
are required to effectively model early SN emission and could be affected by these pulsations. 
Nonetheless, this work highlights the influence of the complete density profile of the progenitor star 
on the SN emission on the plateau, beyond the initial shock cooling and early spherical phase.

\acknowledgements

We would like to thank Evan Bauer for formative conversations and \GYRE\ insights. 
We thank Rich Townsend for discussions about \GYRE.
We would also like to thank 
Matteo Cantiello, 
Rob Farmer, 
Josiah Schwab, 
and Frank Timmes for helpful discussions.
It is a pleasure also to thank 
Maria Drout,
Daichi Hiramatsu,
and
Christopher Kochanek
for discussions and correspondences about observations.
This research benefited from interactions with 
Jim Fuller, 
Adam Jermyn,
David Khatami,
Sterl Phinney,
and
Eliot Quataert
which were funded by the Gordon and Betty Moore Foundation through Grant GBMF5076.

J.A.G. is supported by the National Science Foundation (NSF) Graduate Research Fellowship under 
grant number 1650114. The \MESA\ project is supported by the NSF
under the Software Infrastructure for Sustained Innovation program grant ACI-1663688.
This research was supported 
at the KITP by the NSF under grant PHY-1748958.

This research made extensive use of the SAO/NASA Astrophysics Data System (ADS).

\software{
\texttt{Python} from \href{https://www.python.org}{python.org}, 
\texttt{py\_mesa\_reader} \citep{MesaReader}, 
\texttt{ipython/jupyter} \citep{perez2007ipython,kluyver2016jupyter}, 
\texttt{SciPy} \citep{scipy}, 
\texttt{NumPy} \citep{der_walt_2011_aa}, and 
 \texttt{matplotlib} \citep{hunter_2007_aa}.
}

\renewcommand*{\bibfont}{\small}
\bibliographystyle{apj2}
\bibliography{IIP_CCSNe}


\end{document}